# Random Triggering Based Sub-Nyquist Sampling System for Sparse Multiband Signal

Yijiu Zhao, Yu Hen Hu, and Jingjing Liu

*Abstract*—We propose a novel random triggering based modulated wideband compressive sampling (RT-MWCS) method to facilitate efficient realization of sub-Nyquist rate compressive sampling systems for sparse wideband signals. Under the assumption that the signal is repetitively (not necessarily periodically) triggered, RT-MWCS uses random modulation to obtain measurements of the signal at randomly chosen positions. It uses multiple measurement vector method to estimate the non-zero supports of the signal in the frequency domain. Then, the signal spectrum is solved using least square estimation. The distinct ability of estimating sparse multiband signal is facilitated with the use of level triggering and time to digital converter devices previously used in random equivalent sampling (RES) scheme. Compared to the existing compressive sampling (CS) techniques, such as modulated wideband converter (MWC), RT-MWCS is with simple system architecture and can be implemented with one channel at the cost of more sampling time. Experimental results indicate that, for sparse multiband signal with unknown spectral support, RT-MWCS requires a sampling rate much lower than Nyquist rate, while giving great quality of signal reconstruction.

## I. INTRODUCTION

IN radio frequency (RF) signal processing systems, such as communication and radar systems, the spectrum of wideband RF signals are often populated by few sparsely allocated narrowband spectrums [1], [2]. Such signals are called *sparse multiband signals* [3]. Due to the wideband nature, high-speed analog-to-digital converter (ADC) will be required to capture these sparse multiband signals at Nyquist rate. If the required sampling rate exceeds the specification of available ADC, alternative sampling approaches must be taken to address this challenge.

Multi-coset sampling (MCS) is a non-uniform periodic sampling method [4], [5] for sub-Nyquist rate sampling. MCS consists of a bank of ADCs clocked at the same rate but with different phases to facilitate concurrent sampling of the wideband RF signal at different delays. The accuracy of MCS sample depends on the accuracy of the delay circuitry which is very complex and expensive. For periodic wideband signals sampling, random equivalent sampling (RES) [6], [7] only requires a single ADC clocked at sub-Nyquist rate. It accomplishes this by exploiting the phase incoherence between the sampling clock and the periodic signal. As such, a single cycle waveform of the wideband periodic signal may be reconstructed using time-alignment method, giving the equivalent effect of sampling the signal at Nyquist rate. However, to achieve desirable accuracy, this sampling process requires considerable time to complete. For instrumentation applications, time-interleaved sampling (TIS) [8], [9], which is a special case of MCS, is another widely used sampling technique. Its effective sampling frequency is proportional to the number of ADCs used in the system. Yet, its accuracy suffers when the ADC used is much slower than the Nyquist rate. For non-periodic wideband signals, TIS often is the only viable choice for signal sampling. While they have different pros and cons, the performance of MCS, RES and TIS sampling techniques are all limited by the input bandwidth barrier of ADC [10].

Compressive sampling (CS) [11], [12] has been proposed as an emerging sub-Nyquist rate sampling technique for inherently sparse signals. Based on a random demodulation technique, Kirolos *et al.* [13], [14] developed an analog-to-information converter (AIC) to realize sub-Nyquist rate sampling of wideband signal using CS reconstruction. However, to accomplish this, the wideband signal will be modulated by a pseudorandom pulse sequence at the Nyquist rate. For wideband signals consisting of few scattered harmonics, AIC has been shown an effective sampling method. For sparse multiband signal, Eldar *et al.* [3], [15] developed a modulated wideband converter (MWC) based on the random demodulation technique. In particular, MWC operates multiple AIC samplers concurrently so that available MWC channels is proportional to the number of sparsely allocated narrow bands in the signal spectrum. MWC consumes considerable amount of hardware and the realization of the random modulation pulse sequence is rather complicated. In [16], frequency down-conversion is employed to decrease the number of sampling channels of MWC, and it requires additional preprocessing circuitry in the channels.

This work is supported by the National Natural Science Foundation of China (Grant Nos. 61301264 and 61671114), and by the Ph.D. Programs Foundation of Ministry of Education of China (Grant No. 20130185120019).

Y. Zhao and J. Liu are with the School of Automation Engineering, University of Electronic Science and Technology of China, Chengdu 611731, China (e-mail: yijiuzhao@uestc.edu.cn; liujjing@hotmail.com).

Y. H. Hu is with the Department of Electrical and Computer Engineering, University of Wisconsin-Madison, Madison, WI 53706, USA (e-mail: hu@engr.wisc.edu).

Previously [17], [18], we have incorporated CS theory to enhance efficiency of RES signal reconstruction (called CS-RES) with satisfactory result. We have shown that with CS based reconstruction, much fewer RES samples will be needed to reconstruct the periodic signal and the impact of phase coherence can also be mitigated. CS-RES can be applied to both harmonic sparse signal and sparse multiband signal, the accuracy is still limited by the bandwidth restriction of ADC operating at sub-Nyquist rate.

In this work, we propose a *Random-Triggering based Modulated Wideband Compressive Sampling* (RT-MWCS) method. Specifically, a repetitively excited sparse multiband signal is sampled under control of level triggering circuitry. Once triggered, it is demodulated by a periodic pseudorandom sequence at the Nyquist rate. The output then is low pass filtered and sampled with an ADC clocked at a sub-Nyquist rate. Each excitation yields one sampling sequence. With sufficient number of sampling sequences are obtained, a multiple measurement vector (MMV) method is applied to estimate the non-zero support of the frequency spectrum of the sparse multiband signal, and finally, the frequency spectrum is recovered based on the estimated support and least square estimation.

In comparison to the popular sampling approaches, such as e.g. MCS, RES, and TIS, the proposed RT-MWCS samples the baseband signal, and it is not subject to the input bandwidth barrier of ADC. The main contribution of this paper is to propose a novel random sampling method that incorporates the advantages of both random modulation and random equivalent sampling. As such, by introducing the random triggering technique, potential hardware implementation may be simplified. The numerical simulation and hardware evaluation demonstrate that the proposed RT-MWCS is efficient and robust to noise.

The remaining of this paper is organized as follows. In Section II, the wideband sampling problem is formulated. In Section III, the RT-MWCS system is proposed, and a comparison of RT-MWCS with related work is presented. In Section IV, a hardware implementation is presented. In Section V, both numerical simulation and hardware evaluation results are reported and discussed. Conclusion is summarized in Section VI.

## II. FORMULATION OF THE PROBLEM

A sparse multiband signal is a bandlimited, square-integrable, continuous time signal whose spectrum is zero-valued except a set of disjoint narrow frequency bands where signal energy concentrates [19]. An example of the spectrum of a sparse multiband signal is illustrated in Fig. 1.

Let $x(t)$ be a real valued sparse multiband signal with the Fourier transform [3]

$$X(f) = \int_{-\infty}^{+\infty} x(t)e^{-j2\pi ft} dt . \quad (1)$$

Denote $f_{NYQ}$ to be Nyquist rate sampling frequency of $x(t)$, and $T$

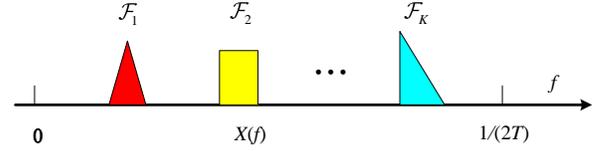

Fig. 1. Spectrum-sparse multiband signal.

$= 1/f_{NYQ}$ to be the sampling period. Let $\mathcal{F}_i$ be the $i^{th}$ narrow frequency band where $X(f) \neq 0$. The band positions are arbitrary, and unknown in advance. The support of the frequency spectrum of a sparse multiband signal $\mathcal{F} \subset [-1/(2T), 1/(2T)]$ may be defined as: $\mathcal{F} = \bigcup_{i=1}^{K} \mathcal{F}_i$, $\mathcal{F}_i \cap \mathcal{F}_j = \varnothing$ (empty set) if $i \neq j$. Note that $X(f) = 0$ if $f \notin \mathcal{F}$. Let $[\mathcal{F}] = [f_{min}, f_{max}]$ denote the spectral span, which is the smallest interval containing $\mathcal{F}$. One may define the spectral occupation ratio $Q = \lambda(\mathcal{F})/|[\mathcal{F}]|$ where $\lambda(\mathcal{F})$ is the Lebesgue measure of $\mathcal{F}$, and $|[\mathcal{F}]| = f_{max} - f_{min}$. Clearly, one has $0 < Q < 1$. Since the spectrum may be shifted to the origin by modulating the time domain signal, one may conveniently set $f_{min} = 0$ and $f_{max} = 1/(2T)$. To be qualified as a sparse multiband signal, one requires $Q \ll 1$.

If the positions of the disjoint sub-bands in a sparse multiband signal are *known*, one may modulate $x(t)$ with a harmonic signal with a frequency at the middle of $\mathcal{F}_i$. Then, the component associated with $\mathcal{F}_i$ can be isolated from the remaining sub-bands by applying a low pass filter (LPF) on the modulated analog signal with passing frequency equal to half bandwidth of $\mathcal{F}_i$. As such, the baseband signal can be acquired using an ADC operating at sub-Nyquist rate. Unfortunately, in practice, the frequency spans of individual $\mathcal{F}_i$ are unknown.

## III. SUB-NYQUIST SAMPLING SYSTEM

### A. System Description

A block diagram of the proposed RT-MWCS system is depicted in Fig. 2. In RT-MWCS, signal is reconstructed from multiple acquisitions. Since the sampling system is based on RT technique, a fixed reference point is required in each acquisition. Generally, periodic signal and repetitively triggered signal satisfy the requirement of RT-MWCS. Moreover, periodic signal could be treated as a special case of repetitively triggered signal. Therefore, we assume the sparse multiband signal $x(t)$ has finite duration and is repetitively triggered. Examples of such signal include [20]-[23]. The rate of repetition need not be constant. Hence $x(t)$ may also be treated as an aperiodic signal. To acquire the waveform of $x(t)$, we will repeatedly trigger the signal $M$ times.

During the $m^{th}$ excitation ($1 \leq m \leq M$), when the value of $x(t)$ exceeds a preset threshold level, the control module will trigger the pseudorandom sequence generator to apply a periodic random modulation signal $p_m(t)$ to modulate $x(t)$ and result in a modulated signal $\tilde{x}_m(t) = x(t)p_m(t)$. This modulated signal then will pass through an analog LPF with output denoted by $y_m(t)$. Next, one digitized sampling sequence will be obtained from the ADC that runs at sub-Nyquist rate.

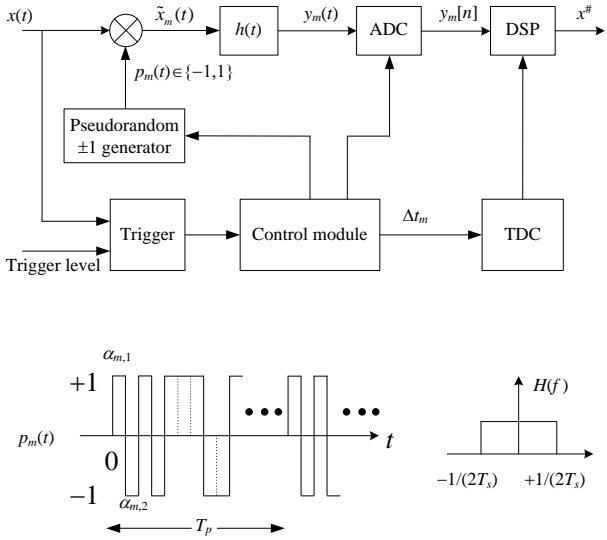

Fig. 2. RT-MWCS block diagram.

To elaborate, assume $x(t)$ exhibits identical waveform during each excitation, if the triggering level remains constant, then the triggering positions at the waveform should be the same. Yet the sub-Nyquist rate clock that controls the ADC will not be in synchrony with the (often irregular) excitation rate. Hence the portion of the waveform on $y_m(t)$ that will be acquired by the ADC will be random.

The pseudorandom sequence $p_m(t)$ is with period of $T_p = 1/f_p$. In each period of $p_m(t)$, there are $L$ pulses each with duration $T_p/L$ and magnitude $\{\alpha_{m,l}\}$ ($\alpha_{m,l} \in \{+1, -1\}$, $1 \le l \le L$) [3].

The LPF shall have a cutoff frequency equal to $f_s/2$ where $f_s = 1/T_s$ is the sampling frequency of the ADC and $T_s$ is the sampling period. In general, it is selected such that

$$f_{NYQ} = L \cdot f_s, \quad \max_{1 \le i \le K} |[\mathcal{F}_i]| \le f_p \le f_s. \qquad (2)$$

The purpose of this LPF is to minimize the aliasing effect. Since the repetitive excitation of $x(t)$ is not synchronized with the sampling clock of the ADC, the time difference between the threshold triggering of the $x(t)$ to the starting clock edge of ADC sampling, denoted by $\Delta t_m$ will be in general a random quantity between 0 and $T_s$. $\Delta t_m$ could be measured using a time-to-digital converter (TDC) circuitry [24]. As such, the corresponding position of the acquired sample from ADC within the duration of $x(t)$ may be determined and used to establish the relation between the unknown signal and the known samples in the signal reconstruction stage.

*B. Sampling Model Analysis*

The pseudorandom modulation signal $p_m(t)$ in the $m^{th}$ acquisition is a periodic signal with period $T_p$. Hence it can be represented as a Fourier series expression:

$$p_m(t) = \sum_{l=-\infty}^{\infty} c_{m,l} e^{j\frac{2\pi}{T_p}lt} \qquad (3)$$

where the Fourier series coefficient,

$$c_{m,l} = \frac{1}{T_p} \int_0^{T_p} p_m(t) e^{-j\frac{2\pi}{T_p}lt} dt. \qquad (4)$$

The Fourier transform of random-modulated signal $\tilde{x}_m(t)$ is the convolution of the Fourier transform of $x(t)$, denoted by $X(f)$, and the Fourier series coefficients of $p_m(t)$:

$$\tilde{X}_m(f) = \int_{-\infty}^{\infty} \tilde{x}_m(t) e^{-j2\pi ft} dt = \sum_{l=-\infty}^{\infty} c_{m,l} X(f - lf_p). \qquad (5)$$

The output of the modulator is a linear combination of $f_p$ shifted copies of $X(f)$. Since $X(f)$ is bandlimited within $[-f_{NYQ}/2, f_{NYQ}/2]$, and the maximum active band width is smaller than $f_p$, the sum in (5) contains no more than $\lceil f_{NYQ}/f_p \rceil$[1] nonzero terms. In this work, the filter $H(f)$ is assumed to be an ideal LPF as illustrated in Fig. 2 (In practical implementation, the designed LPF circuitry is nonideal, and it can be approximated and compensated in the digital domain [25]). Therefore, only frequencies in pass band $\mathcal{F}_s = [-1/(2T_s), 1/(2T_s)]$ are retained in the Fourier transform of $y_m(t)$:

$$Y_m(f) = \tilde{X}_m(f) H(f) = \sum_{l=-L_0}^{L_0} c_{m,l} X(f - lf_p), \quad f \in \mathcal{F}_s. \qquad (6)$$

To simplify the expression we assume $f_p = f_s$. In order to make $Y_m(f)$ contain all nonzero contributions of $X(f)$, $L_0$ is chosen as the smallest integer satisfying $2L_0 + 1 \ge f_{NYQ}/f_p$.

For each acquisition, the phase difference between the trigger pulse and sampling clock is asynchronous, while, relating to the underlying signal, the trigger pulse is fixed. Equivalently, $\Delta t_m$ is the time offset between $y_m(t)$ and sampling clock, and $y_m(t)$ is sampled after left shifted $\Delta t_m$. Consequently, the $m^{th}$ random sampling sequence has following expression:

$$y_m[n] = y_m(n \cdot T_s + \Delta t_m), \quad n \in \mathbb{Z} \qquad (7)$$

and its discrete-time Fourier transform (DTFT) is

$$\begin{aligned} Y_m(e^{j2\pi fT_s}) &= \sum_{n=-\infty}^{\infty} y_m[n] e^{-j2\pi fnT_s} \\ &= \int_{-\infty}^{\infty} y_m(t + \Delta t_m) e^{-j2\pi ft} dt = Y_m(f) e^{j2\pi f\Delta t_m} \end{aligned}, \quad f \in \mathcal{F}_s. \qquad (8)$$

---

[1] $\lceil x \rceil$ denotes the ceiling operator, which returns the smallest integer not less than $x$.

Combining (6) and (8), one has

$$e^{-j2\pi f \Delta t_m} Y_m(e^{j2\pi fT_s}) = \sum_{l=-L_0}^{L_0} c_{m,l} X(f - lf_p), f \in \mathcal{F}_s. \quad (9)$$

Eqn. (9) ties the DTFTs of known ADC measurements $y_m[n]$ to the unknown signal $X(f)$.

Denote

$$\phi_{m,l} = c_{m,-l}, 1 \leq m \leq M, \quad (10a)$$

$$s_l(f) = X\left(f + (l - L_0 - 1)f_p\right), 1 \leq l \leq L = 2L_0 + 1, \quad (10b)$$

and

$$z_m(f) = e^{-j2\pi f \Delta t_m} Y_m(e^{j2\pi fT_s}). \quad (10c)$$

In (10a), the reverse order is due to the enumeration of $s_l(f)$ in (10b). Then (9) may be represented in a matrix-vector formulation:

$$\begin{bmatrix} z_1(f) \\ z_2(f) \\ \vdots \\ z_M(f) \end{bmatrix} = \begin{bmatrix} \phi_{1,1} & \phi_{1,2} & \cdots & \phi_{1,L} \\ \phi_{2,1} & \phi_{2,2} & \cdots & \phi_{2,L} \\ \vdots & \vdots & \ddots & \vdots \\ \phi_{M,1} & \phi_{M,2} & \cdots & \phi_{M,L} \end{bmatrix} \begin{bmatrix} s_1(f) \\ s_2(f) \\ \vdots \\ s_L(f) \end{bmatrix}$$

$$\Leftrightarrow \mathbf{z}(f) = \mathbf{\Phi} \mathbf{s}(f), \quad f \in \mathcal{F}_s. \quad (11)$$

Fig. 3 depicts the relation between $\mathbf{s}(f)$ and $X(f)$ with $K = 2$ pairs of sub-bands.

To reconstruct signal $x(t)$, we need to find a unique solution for (11). Eqn. (11) is constructed in the frequency domain. Therefore, signal reconstruction process could be divided into two stages: support estimation in the frequency domain and signal reconstruction in the time domain.

In RT-MWCS model, the entries of measurement matrix $\mathbf{\Phi}$ are Fourier series coefficients of pseudorandom sequences. Therefore, rank($\mathbf{\Phi}$) = min{$M, L$}, and there are two cases to solve (11).

Case 1: $M \geq L$. Eqn. (11) is an over-determined linear system of equations and $\mathbf{s}(f)$ may be solved as the least square solution. However, large $M$ implies more rounds of sampling acquisitions, and hence longer measurement time to produce a sufficiently accurate reconstruction. Moreover, according to (2), smaller $L$ implies higher ADC sampling rate $f_s$.

Case 2: $M < L$. Eqn. (11) is an under-determined linear system of equations and may have infinite many solutions. However, being a sparse multiband signal, the vector $\mathbf{s}(f)$ inherently will have a sparse structure that may be exploited using compressive sensing reconstruction [26] to estimate the non-zero entries (supports) of the vector $\mathbf{s}(f)$.

The task of *support estimation* is to identify which elements of $\mathbf{s}(f)$ contain active bands. Since $X(f)$ is partitioned into $L$ equal width frequency slices that comprise $\mathbf{s}(f)$, identifying the nonzero entries of $\mathbf{s}(f)$ can only approximate the true frequency band of $x(t)$ upto a resolution of $1/(LT)$ Hz. Similar to spectrum sensing using MCS [27], support recovery could be realized by examining the covariance matrix of the random sampling sequences.

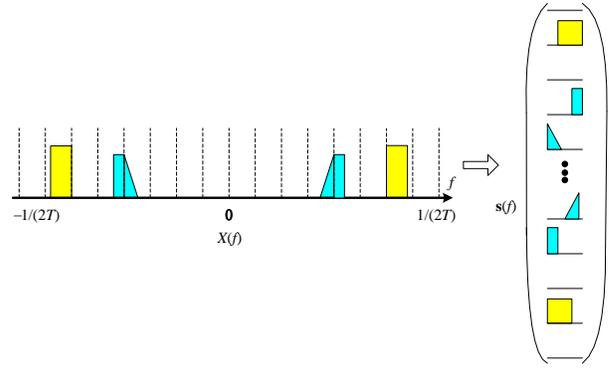

Fig. 3. $X(f)$ is partitioned into $L$ equal width frequency slices that are the entries of vector s($f$).

Define the $M \times M$ covariance matrix of $\mathbf{z}(f)$

$$\mathbf{R} := \int_{-1/(2LT)}^{1/(2LT)} \mathbf{z}(f)\mathbf{z}(f)^H \, df. \quad (12)$$

$\mathbf{R}$ cannot be evaluated directly. Instead, using the Parseval theorem,

$$\mathbf{R}_{i,k} = \langle z_i[n], z_k[n] \rangle = \sum_{n=-\infty}^{\infty} z_i[n] \cdot z_k^*[n]. \quad (13)$$

where $\{z_i[n]; -\infty < n < \infty\}$ is the time domain Fourier series corresponding to $\mathbf{z}(f)$. From (10c), $z_m[n]$ is the time-shifted sequence of output sample $y_m[n]$. However, the amount of desired time shift $\Delta t_m$ is *not* an integral multiple of the ADC sampling time $T_s$. To resolve this difficulty, we choose to up-sample (interpolate) $y_m[n]$ $L$ times by inserting $L-1$ zeros between successive elements, and then passing the sequence through a LPF with cutoff frequency $f_s/2$. The resulting up-sampled sequence, denoted by $\tilde{y}_m[n]$, will now have a sampling rate of $f_{NYQ}$ (= $L \cdot f_s$). Finally, denote

$$\tau_m = \lceil \Delta t_m / T \rceil = \lceil \Delta t_m \cdot f_{NYQ} \rceil \quad (14)$$

then,

$$z_m[n] = \tilde{y}_m[n - \tau_m]. \quad (15)$$

With $\{z_m[n]\}$ estimated, one may proceed to compute $\mathbf{R}$ in (13). Once $\mathbf{R}$ is obtained, CS idea is employed to directly estimate the support of $\mathbf{s}(f)$ [26], and the time domain signal could be reconstructed based on inverse Fourier transform.

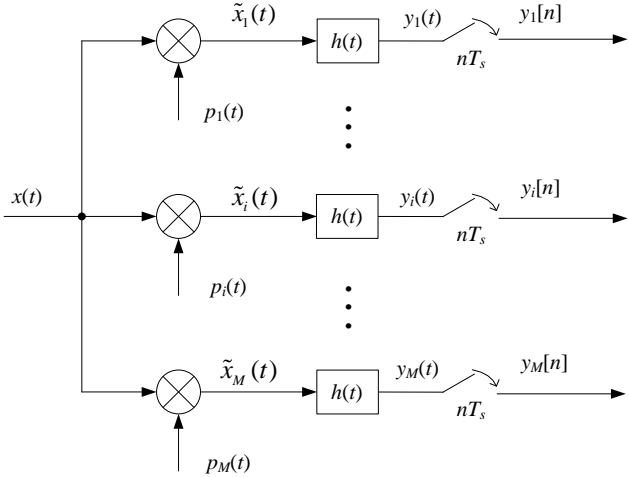

Fig. 4. Block diagram of MWC. MWC contains of multiple channels.

TABLE I
COMPARISON BETWEEN MWC AND RT-MWCS

| Sampling approach | **MWC** | **RT-MWCS** |
|---|---|---|
| Work for sparse multiband signal | Yes | Yes |
| Subject to ADC bandwidth limitation | No | No |
| Reconstruction performance | Good | Good |
| Time consumption | Less | More |
| Power consumption | More | Less |
| Architecture complexity | Complex | Simple |
| Implementation area | Large | Small |

*C. Comparison to Related Work*

MWC is also a sub-Nyquist sampling system based on random demodulation technique, as shown in Fig. 4 [3]. Signal is fed into a bank of modulators and demodulated by pseudorandom sign waveforms to alias the spectrum into baseband. Then, the demodulated signals are low pass filtered and uniformly sampled using a bank of ADCs clocked at low rate. Finally, the signal is reconstructed by solving MMV problem. MWC system could be used to sample the sparse multiband signal at sub-Nyquist rate. However, in order to stably reconstruct signal, MWC requires roughly [3]

$$M \approx 8K \log(L/4K) \quad (16)$$

channels to estimate the signal support while RT-MWCS requires only one channel. Moreover, with the presence of multiple channels, synchronizing $M$ ADC clocks also presents significant implementation challenges.

Despite the somewhat similar mathematical analysis of MWC and RT-MWCS, their system architectures are quite different. In MWC, multiple channels synchronously sample the demodulated signal, and signal is reconstructed from sampling sequences captured in the same acquisition run. While, under control of trigger pulse, our proposed RT-MWCS uses a single channel to sample signal. In order to collect enough information of signal, multiple acquisition runs are required in RT-MWCS, and the number of acquisition runs needs to satisfy (16). Consequently, in comparison to MWC, RT-MWCS requires more sampling time. For each acquisition run, signal is demodulated by a different pseudorandom sign waveform. Obviously, RT-MWCS has much simpler architecture.

Table I presents the comparison between MWC and RT-MWCS.

## IV. HARDWARE IMPLEMENTATION

The proposed RT-MWCS was implemented in hardware module, which consists of two modules: a sampling module and a reconstruction module, as shown in the block diagram in Fig. 5. The sampling module processes the input analog signal and acquires the sub-Nyquist samples. In the reconstruction module, the obtained digital sample data are transmitted through universal serial bus (USB) to a personal computer (PC). Then, signal is reconstructed using Matlab algorithm that is performed offline.

In the sampling module, the input signal is mixed with a high speed pseudorandom binary sequence. In order to sense the underlying signal in the frequency domain, the pseudorandom sequence should be clocked at a rate that is no lower than the signal Nyquist rate. This rate is also the equivalent sampling rate of the reconstructed signal. It is a challenge to generate high speed sequence. Considering configurability of sequence, we propose a field-programmable gate array (FPGA) (EP4CE10F256, Altera) device based pseudorandom binary sequence generation module. Due to speed limitation of FPGA, parallel to serial converter (MC100EP446, ON Semiconductor) is employed in hardware implementation. In CS based sampling techniques, the measurement matrices constructed by the pseudorandom sequences should satisfy the restricted isometry property (RIP). In particular, it has been proved that a measurement matrix with independent and identically distributed Bernoulli entries meets the RIP condition [28]. Therefore, the encoded parallel data with Bernoulli distribution are generated using Matlab and stored in the read only memory (ROM) of FPGA. Considering multiple acquisition runs, there should be enough pseudorandom sequences stored in the ROM(In this implementation, 100 sequences are stored in the ROM, and each sequence is with length of 197). The converter converts 8-bit width parallel data into a bit stream. In this implementation, the equivalent sampling rate is set to 2.5 GHz. Consequently, the converter operates at a rate of 2.5 GHz, and the parallel data output from FPGA is with rate of 312.5 MHz that is within the speed capacity of FPGA pins. In order to obtain a sequence $p_m(t)$ as depicted in Fig. 2, the bit stream is alternating current (AC) coupled and conditioned.

The output of mixer (SYM-30DHW, Min-Circuits) is low pass filtered before sampling. The filtered signal is sampled using a low speed ADC (AD9648, Analog) under control of trigger pulse. In this work, the trigger pulse is generated using a comparator (ADCMP562, Analog). The signal is compared with a level that is set using a digital to analog converter (DAC) (AD5322, Analog). In RT-MWCS, the trigger pulse and sampling clock are asynchronous. The shift time $\Delta t$ between the trigger pulse to sampling clock provides a relative position in signal reconstruction and is required to be measured. We measure $\Delta t$ using a TDC module, which is implemented with fast charging and slow discharging circuits. In this paper, TDC stretches the $\Delta t$ (narrow pulse). Then, the output of TDC is fed into FPGA, and counted using a high speed clock [17].

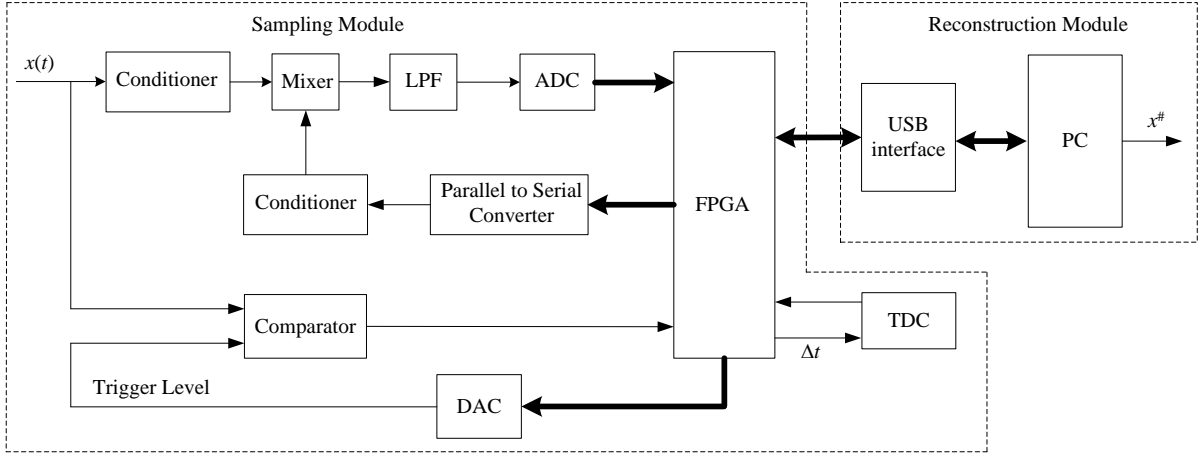

Fig. 5. Block diagram of RT-MWCS implemented in hardware module.

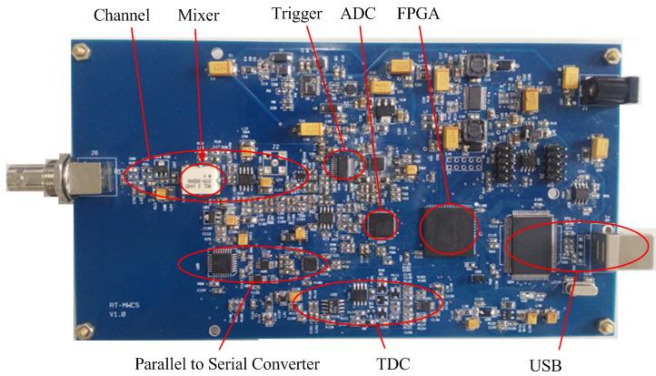

Fig. 6. Picture of the RT-MWCS prototype.

Once the sampling sequences are obtained, they are transmitted to PC through USB interface device (CY7C68013, Cypress). The USB interface device integrates an "8051" micro processor unit, in which the acquisition software is operated. Picture of the RT-MWCS prototype is shown in Fig.6.

In hardware implementation of RT-MWCS, the designed LPF is nonideal, and it will introduce mismatch between circuit implementation and (6). The designed LPF can be calibrated at the cost of sampling frequency [25]. On the other hand, the stretch ratio of TDC may be affected by the temperature. To measure the time difference accurately, TDC needs to be calibrated using the preset pulses [29].

## V. Experiment

In this section, numerical simulation and hardware evaluation are reported to investigate the proposed RT-MWCS.

### A. Numerical Simulation

To evaluate the performance of the proposed system (see Fig. 2), numerical experiments with a test signal are performed. An interesting application of RT-MWCS is sampling and reconstruction of sparse multiband signal, whose active band locations are unknown. In the experiments, the test signal is defined as following

$$x(t) = \sum_{i=1}^{K} \sqrt{E_i B} \cdot \text{sinc}\big(B(t-t_i)\big)\cos\big(2\pi f_i(t-t_i)\big) \quad (17)$$

where $K$ is the number of pairs of active bands, $E_i$ is the energy coefficient, $B$ is the sub-band bandwidth, $t_i$ is the time offset with respect to $t = 0$, and $f_i$ is the carrier frequency. In all experiments, $B = 10$ MHz, $t_i$ is randomly chosen in $[0, 10]$ $\mu$s, and $E_i$ is randomly chosen in $[1, 10]$. If $f_i + B/2 \leq f_{NYQ}/2$, then the signal may be equivalently sampled without aliasing. Therefore, $f_i$ is randomly chosen in $[B/2, (f_{NYQ} - B)/2]$. The equivalent sampling rate of reconstructed signal is $f_{NYQ} = 2.5$ GHz.

In the sampling stage, $f_s$ and $f_p$ satisfy the following relation

$$f_s = f_p = \frac{f_{NYQ}}{L} \geq B . \quad (18)$$

In each acquisition, signal is first mixed with a pseudorandom sequence, which is with equivalent sampling rate of $f_{NYQ}$. In the following experiments, the sampling rate is chosen as $f_s = f_p = f_{NYQ}/197 \approx 12.7$ MHz, which meets the requirement of (18) and is much lower than the Nyquist rate of $f_{NYQ}$. A sampling sequence is randomly captured, and the time interval $\Delta t_m$ is quantized into an integral multiple of $T$. This integral value will be used as the delay in the equivalent Nyquist samples. After collecting enough random sampling sequences, model (11) is constructed based on time intervals and pseudorandom sequences. And then, based on (11), signal is reconstructed with CS algorithm.

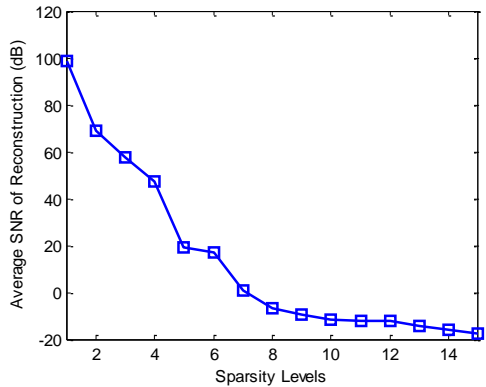

Fig. 7. Reconstruction performance with respect to different sparsity level.

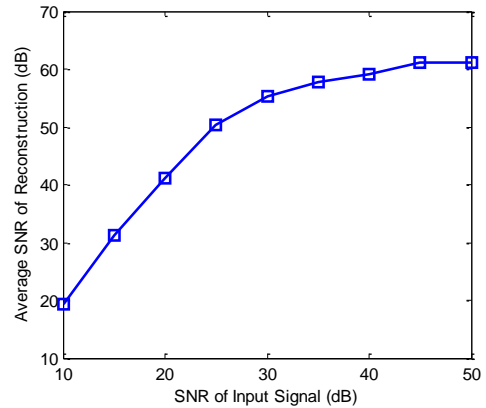

Fig. 8. Reconstruction performance with respect to different SNR of input signal.

In general, more sampling sequences are required to reconstruct a less sparse signal with a desired accuracy. In the first experiment, we investigate the reconstruction performance with respect to the sparsity level $K$. Sparsity levels over range of 1 to 15 in increment of 1 are investigated. For each specific sparsity level, 200 random trials are performed. In this experiment, $M = 20$ sampling sequences are considered in each reconstruction. The average signal-to-noise ratio (SNR) of reconstruction is depicted in Fig. 7. Obviously, larger sparsity level degrades the signal reconstruction accuracy, and more sampling sequences are required. Fortunately, more sampling sequences (larger $M$) means longer sampling time, and it does not complicate the system architecture.

In practical application, the signal may be corrupted by noise, or the noise may be introduced in the sampling stage. In this experiment, we consider more practical situation that the underlying signal is corrupted by noise. In the sampling stage, the white Gaussian noise is added in the test signal, and signal parameters ($K = 3$) are fixed in all trials. $M = 20$ random sampling sequences are used to reconstruct signal. The input signal with SNR over the range of 10 to 50 dB in increment of 5 dB are tested. 200 random trials are performed for each specific SNR value, and the averaged output SNR is shown in Fig. 8. Clearly, the proposed compressive sampling system is robust against additive white Gaussian noise.

The last simulation is performed to compare the performance of the proposed RT-MWCS against the existing CS-based sub-Nyquist rate sampling approach. As we know, MWC also works for the sparse multiband signal. In this simulation, we consider the reconstruction performance with respect to the number of sampling sequences. For range of 10 to 20 sampling sequences in increment of 1 sequence, 200 random trials are performed for each specific number. The test signal is the same as that of the previous simulation. The reconstructed SNR values averaged over 200 trials with different numbers of sampling sequences are shown in Fig. 9. For both MWC and RT-MWCS, the reconstruction performance is improved with the increase of sampling sequence number. In the experimental signal, sparsity level is $K = 3$. It is clear that after the number of sampling sequences increases beyond 12, the reconstruction yields SNR about 20 dB. Obviously, the MWC reconstruction achieves a bit higher SNR in most of sampling sequence numbers. However, for MWC, more sampling sequences means that more sampling channels are required (one channel takes one sampling sequence). This would be a great challenge in circuit implementation. While, for RT-MWCS, multiple acquisition runs need to be performed, and we do not need to change the system architecture to capture more sampling sequences.

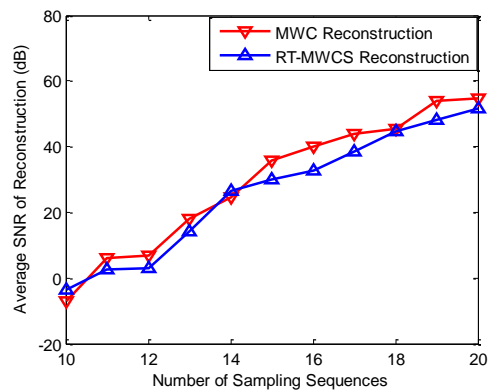

Fig. 9. Comparison between RT-MWCS with MWC.

### B. Hardware Evaluation

The waveform data of noise-free signal (defined in (17)) is generated using Matlab and transmitted to an arbitrary waveform generator (AWG7082C, Tektronix), and it is with sampling rate of 2.5GHz. The synthetic test signal is fed into signal channel of RT-MWCS. For each sample acquisition run, a new pseudorandom sequence is generated and mixed with test signal. Different from numerical simulation, in hardware implementation, the LPF is nonideal. In order to compensate the stop band that is with long tail, sampling sequence is captured at rate of $f_s = 2f_p$ and filtered by the compensation filter that is constructed according to [25]. Before reconstruction, sampling sequence is decimated, and the sampling rate is reduced by factor of 2 to match (11).

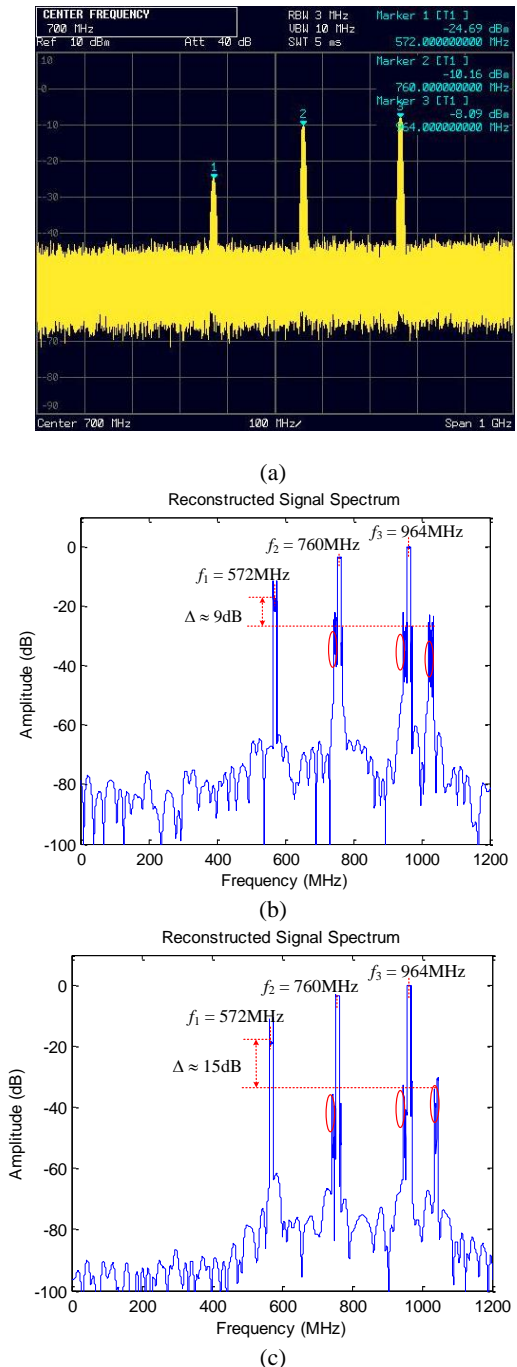

Fig. 10. Comparison between reference and RT-MWCS reconstruction. (a) The spectrum analyzer measurements. (b) Reconstruction from 20 sampling sequences. (c) Reconstruction from 40 sampling sequences.

Fig. 10 shows the comparison between the spectrum analyzer (ESPI7, Rohde&Schwarz) measurements and the proposed RT-MWCS reconstructions. Fig. 10(b) and Fig. 10(c) depict the reconstructions from $M = 20$ and $M = 40$ sampling sequences, respectively. Note that the three carrier frequencies (572MHz, 760MHz and 964MHz) are successfully reconstructed. Due to the nonideal amplitude-frequency response of the designed channel, there are some differences in the signal amplitude values between the measured reference and the reconstruction. In the reconstruction stage, we recover 6 sub-bands with maximum energy. It is clear that there are 3 unwanted sub-bands exist in the reconstruction, which are marked by ellipses. These 3 unwanted sub-bands are introduced by the noise. Since only a part of noise bands are recovered, they look like the sub-bands of signal. Note that the amplitude values of unwanted sub-bands are minimized with the increase of sampling sequences. The reconstructions achieve SNRs of 21.5 dB ($M = 20$) and 24.7 dB ($M = 40$), respectively.

## VI. CONCLUSION

We present a random triggering based modulated wideband signal compressive sampling system, which can be used to sample a sparse multiband signal at sub-Nyquist rate. Our main contributions describe the architecture of compressive sampling system that can be implemented in circuitry, and give the concrete mathematical model that is used to reconstruct signal. The blind spectrum recovery is accomplished by an MMV problem in the framework of CS theory. Compared with the existing sub-Nyquist sampling models, the proposed RT-MWCS model not only achieves low sampling rate but also is with simple system architecture. In RT-MWCS, the reconstruction accuracy can be improved with more random sampling sequences, which is at the cost of sampling time, not complexity of system. A hardware implementation of proposed RT-MWCS is presented. For sparse multiband signal with unknown spectral support, we have tested this sampling model using random sampling sequences with satisfactory results.

In hardware implementation, the time difference of each sequence is generated in FPGA, the uncertainty may exist in the time difference measurement, and the sampling technique based on random triggering method suffers from uncertainty of time difference. Future work includes a comprehensive analysis of the impact of uncertainty of time difference on the reconstructed signal. We wish to construct a relational model between reconstruction and uncertainty of time difference. As a kind of sequential compressive sampling approaches, we also plan to design a halting criterion to adaptively estimate number of active bands with minimum acquisition runs.